\documentclass[journal=jacsat,manuscript=article]{achemso}

\usepackage[version=3]{mhchem} 

\author{Yueqiang Hu}
\altaffiliation{Advanced Manufacturing Laboratory of Micro-Nano Optical Devices, Shenzhen Research Institute, Hunan University, Shenzhen, 518000, China}
\affiliation{National Research Center for High-Efficiency Grinding, College of Mechanical and Vehicle Engineering, Hunan University, Changsha 410082, P.R. China}
\author{$^{,\P}$ Yi Zhang}
\affiliation{National Research Center for High-Efficiency Grinding, College of Mechanical and Vehicle Engineering, Hunan University, Changsha 410082, P.R. China}
\author{Yuting Jiang}
\affiliation{National Research Center for High-Efficiency Grinding, College of Mechanical and Vehicle Engineering, Hunan University, Changsha 410082, P.R. China}
\author{Quan Wang}
\affiliation{National Research Center for High-Efficiency Grinding, College of Mechanical and Vehicle Engineering, Hunan University, Changsha 410082, P.R. China}
\author{Meiyan Pan}
\affiliation{Ji Hua Laboratory, Foshan, 528200, China}
\email{panmy@jihualab.ac.cn}
\author{Huigao Duan}
\affiliation{National Research Center for High-Efficiency Grinding, College of Mechanical and Vehicle Engineering, Hunan University, Changsha 410082, P.R. China}
\altaffiliation{Advanced Manufacturing Laboratory of Micro-Nano Optical Devices, Shenzhen Research Institute, Hunan University, Shenzhen, 518000, China}
\alsoaffiliation{Greater Bay Area Institute for Innovation, Hunan University, Guangzhou,511300, Guangdong Province, China}
\email{duanhg@hnu.edu.cn}
\title{Achromatic Full Stokes Polarimetry Metasurface for Full-color Polarization Imaging in the Visible}

\begin{document}







\begin{abstract} 
\noindent Metasurfaces composed of anisotropic subwavelength structures provide an ultrathin platform for a compact, real-time polarimeter. However, applications in polychromatic scenes are restricted by the limited operating bandwidths and degraded imaging quality due to the loss of spectral information. Here, we demonstrated full-color polarization imaging based on an achromatic polarimeter consisting of four polarization-dependent metalenses. Boosted by an intelligent design scheme, arbitrary phase compensation and multi-objective matching are effectively compatible with a limited database. Broadband achromaticity for wavelengths ranging from 450 nm to 650 nm, with a relative bandwidth of nearly 0.435, is achieved for the full Stokes imaging. The experimental polarization reconstructed errors for operating wavelengths of 450 nm, 550 nm, and 650 nm are $7.5\%$, $5.9\%$, and $3.8\%$, respectively. The full-color and full-polarization imaging capability of the device is also verified with a customized object. The proposed scheme paves the way for further developing polarization imaging toward practical applications.\end{abstract}
\section{Introduction}
\begin{figure}[H]
\centering
\includegraphics[width=\textwidth]{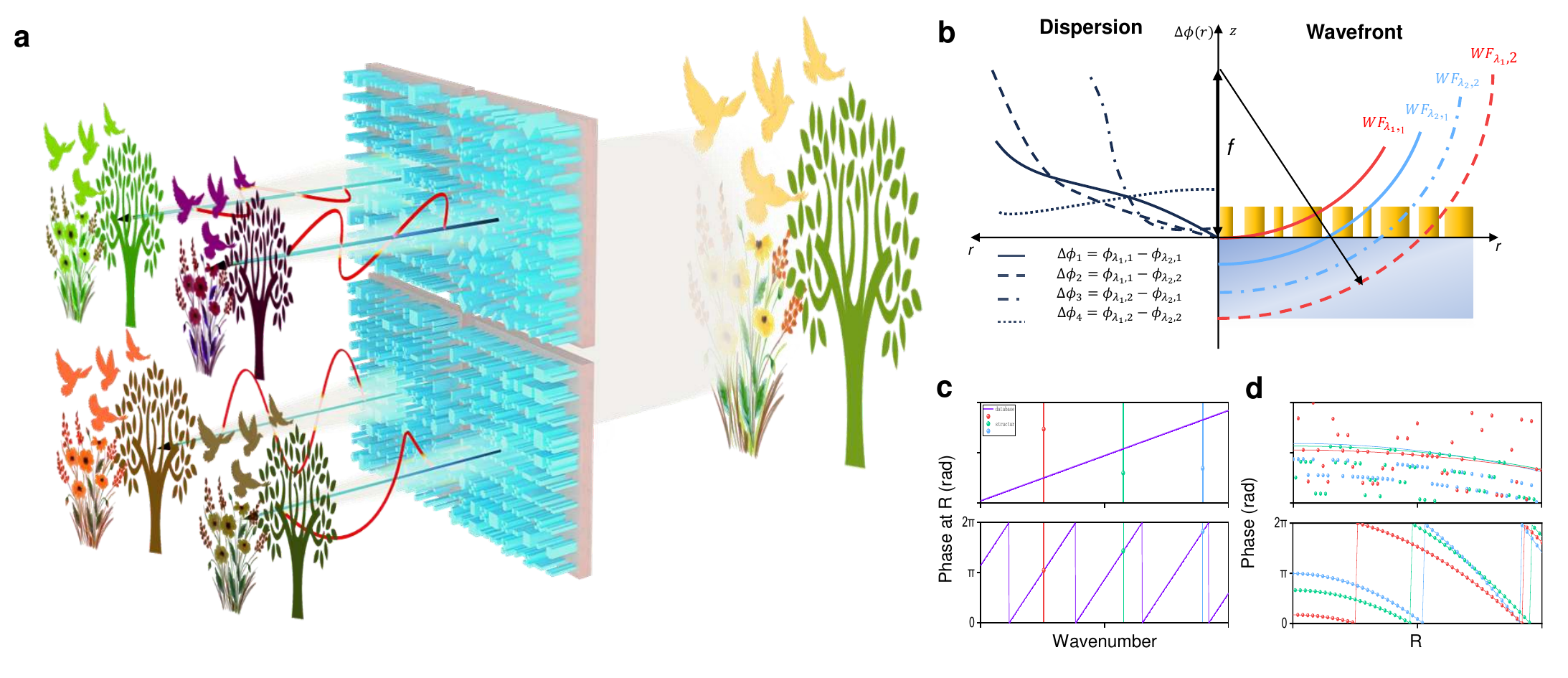}
\caption{ Design of the achromatic imaging polarimeter in the visible. (a) The schematic of the device, which is composed of four sub-metalens that each act as an achromatic focusing lens under a specific polarization. The broadband achromaticity in the 450-650 nm band enables a single shot of the spatial color and polarization information. (b) Arbitrary dispersion is constructed by the wavelength-dependent reference position. (c) Continuous dispersion (upper) and the folded counterpart (bottom). The phases for three wavelengths are represented by dots in red, green, and blue, respectively. The dots represent the realized phases, while the lines are the required phases. The three colors correspond to three different wavelengths, respectively. (d) un-optimized (upper) and optimized (bottom) phase maps. The dots represent the realized phases, while the lines are the required phases. The three colors correspond to three different wavelengths, respectively.}
\label{fig1}
\end{figure}

The full Stokes polarimetry, which can reveal the real-time polarization information of objects, is widely used in applications such as remote sensing\cite{Remote}, materials analysis\cite{Nan09}, and biomedicine\cite{deBoer97}. However, a conventional optical setup for real-time polarization detection consists of multiple optical paths and cascading components, leading to a bulky and complicated system\cite{IJEM2022}. A simple, compact, real-time polarimeter is increasingly required to meet the trending demand in miniaturized optical systems.\\
\indent Metasurface, an ultrathin platform with artificial subwavelength resonators, opens up a new path for designing high-performance planar optical elements such as beam deflectors\cite{RN2394}, metalenses\cite{RN2395,RN2342}, metaholograms\cite{RN2396}, polaroids\cite{Pan2018}, and quantum devices\cite{RN1169}. Particularly, polarization-dependent metasurfaces have been extensively studied for their ability to achieve polarization multiplexing\cite{RN2405} and complex vector beam\cite{RN2326}. Among them, polarization multiplexing metalenses can simultaneously focus basic polarization components of light waves to different spots, offering a new platform for real-time polarization detection and imaging. Many efforts have been made in polarization detection or imaging with metalenses in the visible\cite{RN1101,RN2401,RN1108,RN1115}, near-infrared\cite{RN1106,RN1116,RN2402}, and mid-infrared\cite{RN1107} by either interleaved-unit or sub-metalens schemes. However, these metalenses are designed for a single operation wavelength, while numerous objects contain multi-wavelength information. Hence, the imaging quality for colorful real objects is severely degraded by chromatic aberration and the loss of spectral information. Meanwhile, achromatic metalenses realized through spatial multiplexing\cite{RN1240}, cascaded metasurfaces\cite{RN1239}, or dispersive phase compensation\cite{RN1312,RN1133,RN2403,RN2404,PRL2023,RN2420} have aroused great interest. Recently, broadband polarization detection in the infrared band has been experimentally demonstrated using a long focal depth of focus\cite{RN1254} or dispersion engineering\cite{RN2398}. However, the relative achromatic bandwidth in these works is restricted as the common linear structural dispersion approximation is valid in a limited bandwidth. Expanding the structural dispersion library with more intricate configurations is conventionally required to extend the achromatic broadband, albeit resulting in design and fabrication complexities. An alternative approach is to characterize the phase compensation of meta-atoms with arbitrary nonlinear dispersion\cite{RN2420}. However, achieving efficient matching in multi-objective optimization, which encompasses specific responses to each polarization along with arbitrary phase and dispersion control, becomes increasingly challenging due to the increased degrees of freedom involved.\\
\indent In this work, we demonstrate a broadband achromatic full Stokes metalens designed for full-color polarization imaging within the visible range. As shown in Figure 1a, the device contains four achromatic sub-metalenses, each of which focuses one particular polarization (x-polarization, y-polarization, $45^\circ$-polarization and left circular polarization) to its particular point on the focal plane. The strategy is preferred over the interleaved candidate as the latter suffers from either degraded resolution\cite{RN2399} or restricted size\cite{RN1182}. To extend the design flexibility, arbitrary structural dispersion depending on the spectral reference position is considered.As a means to alleviate the challenges in achieving efficient matching, we manipulate folded dispersions (the bottom figure in Figure \ref{fig1} c) instead of continuous counterparts(the upper figure in Figure \ref{fig1} c). This approach deviates from the conventional achromatic design schemes that rely on continuous dispersions, providing a more manageable solution. Regarding that the utilization of conventional matching by table look-up method is inefficient and difficult for our constructed dispersion, an intelligent matching scheme is developed based on the particle swarm optimization (PSO) algorithm. The complicated matching, which is a multi-objective task, is then efficiently accomplished based on a limited database that does not fill the required phase space (Figure \ref{fig1}d). Consequently, broadband achromaticity for wavelengths ranging from 450 nm to 700 nm, with a relative bandwidth of nearly 0.435, is achieved for the full Stokes imaging. Moreover, the two-step optimization scheme (for orthogonal polarization states) promises the high efficiency of the target polarization and reduces the crosstalk by the orthogonal polarization state simultaneously. The full-color full-Stokes imaging is demonstrated using a self-built polarization mask carrying both color and polarization information. By applying the achromatic polarimetry metasurface to a full-color camera in the visible, the color and the polarization information can be simultaneously obtained in one shot.
\section{Principles}
\begin{figure}[H]
\centering
\includegraphics[width=\textwidth]{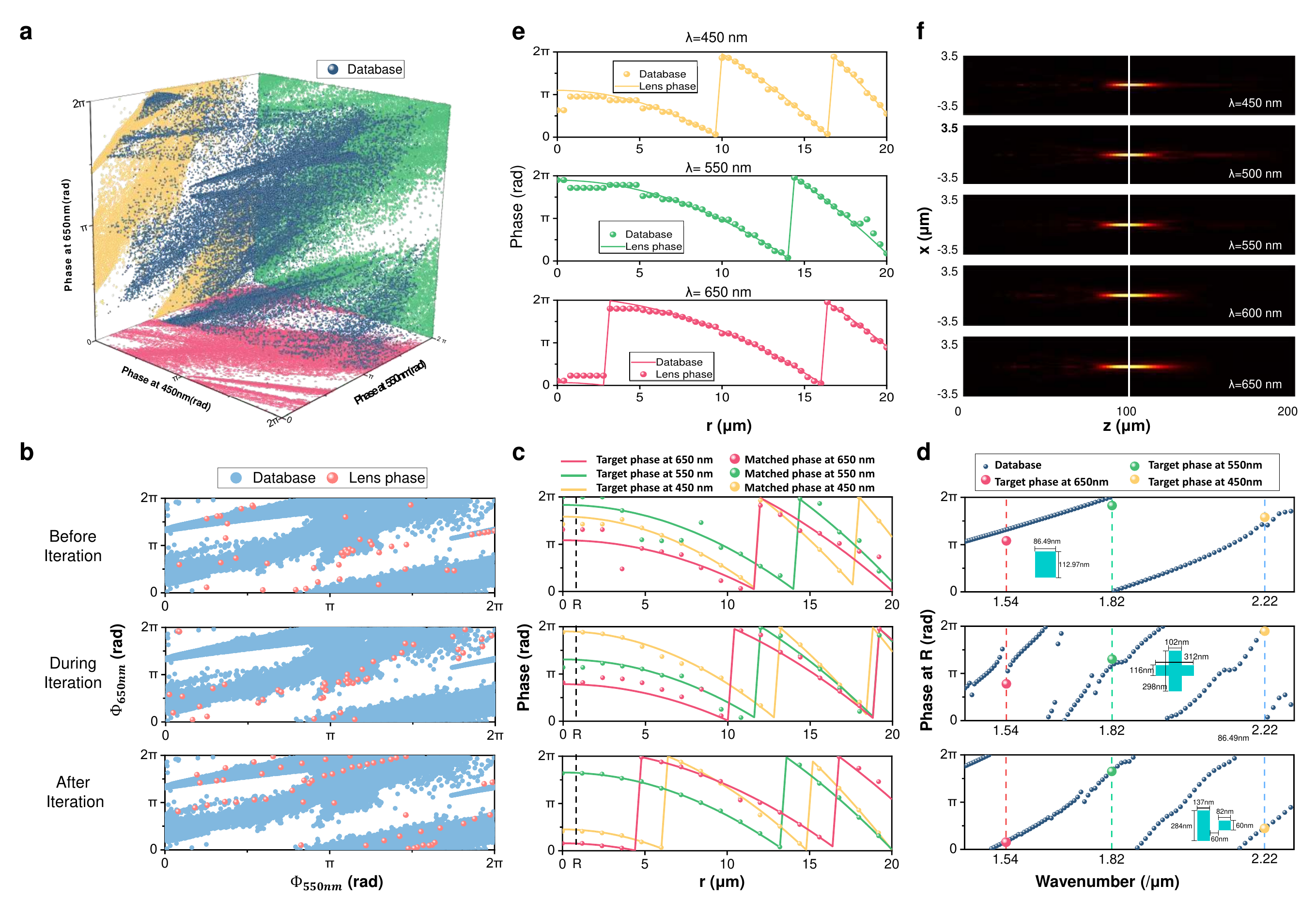}
\caption{(a) Structural dispersion database in the phase space constructed by wavelengths of 450 nm, 550 nm, and 650 nm. (b) Structural dispersion database (blue dots)in the phase plane constructed by wavelengths of 550 nm and 650 nm, and the required lens phases (orange dots) before (top), during (middle), and after (bottom) iterations. (c) Phase matching before (top), during (middle), and after (bottom) iterations. Phases of meta-atoms for the wavelength of 450 nm, 550 nm, and 650 nm are depicted as yellow, green, and red dots, respectively. The corresponding required phase maps are shown as the lines with the same colors. (d) Structural dispersion (blue dots) of a selected meta-atom at the location R depicted in Panel (c). The required phases for the wavelength of 450 nm, 550 nm, and 650 nm are depicted as yellow, green, and red dots, respectively.}
\label{fig2}
\end{figure}

Measuring the image intensities of four necessary polarization components allows recovery of the full Stokes parameter ($S = [S_0, S_1, S_2, S_3]^\mathsf{T}$) (Supplementary Section S5\ref{SI5}) for a complete characterization of the object's polarization information such as the degree of polarization (DOP) and azimuth\cite{RN1108}.\\
\indent For each sub-metalens, the required phase map for the target polarization can be designed as a hyperbolic function:
\begin{equation}
	\Phi_{\parallel} (r,\lambda)=-\frac{2\pi}{\lambda}\left(\sqrt{r^2+f_{t}^2}-f_{t}\right )+C_{0}(\lambda)
\end{equation}
where $r$ is the radial position relative to the center of the sub-metalens; $f_{t}$ is the target focal length; $\lambda_{d}$ is the design wavelength; $C_{0} (\lambda)$ is the reference phase. To improve the matching accuracy, a wavelength-dependent $C_{0} (\lambda)$ corresponding to nonlinear dispersion is constructed\cite{RN2420}:
\begin{equation}
	C_{0}(\lambda)=\frac{2\pi}{\lambda}\left[\sqrt{r_{0}^2(\lambda)+f_t^2}-f_t\right]
\end{equation}
where $r_0 (\lambda)$ is the wavelength-dependent position offering the reference phase. By adjusting $r_0$ at each wavelength, arbitrary complex dispersion can be constructed, and suitable meta-atoms are selected to meet the phase requirements at these discrete wavelengths. That is, the reference position $r_0 (\lambda)$ acts as an optimization factor uniquely determined by the wavelength. \\
\indent To achieve direct full Stokes polarimetry, the wavefront for the orthogonal polarization $\Phi_{\perp} (r,\lambda)$should be manipulated simultaneously to suppress the same focus. In the linear polarization case, the response of the orthogonal polarization is set another focusing phase whose focal length is far away from the target polarization counterpart, while the divergent phase map is naturally achieved for the orthogonal polarization in the circular polarization case as PB phase is utilized (Supplementary Section S1). Therefore, the optimized metalens phase should satisfy two phase maps simultaneously:$\Phi_{\parallel} (r,\lambda)$ for the target polarization, and $\Phi_{\perp} (r,\lambda)$  for the orthogonal polarization. Consequently, the matching is a complicated process using the conventional look-up-table method as multiple objects should be satisfied. To address the issue, two-step matching is proposed. The database is first shrunken with a relatively small wavefront error ($\Delta\Phi_{\parallel}$) for the target polarization:
\begin{equation}
\Delta\Phi_{\parallel}=\sum_{i}\sum_{n}|\Phi_{\parallel} (r_i,\lambda_n )-\Phi_{meta,\parallel}(r_i,\lambda_n )| 
\end{equation}
where $\Phi_{meta,\parallel}(r_i,\lambda_n)$ is the real phase for the target polarization That is, a smaller database is re-constructed with units satisfying $\Delta\Phi_{\parallel}\leq \min{\Delta\Phi_{\parallel}}+\Delta\Phi_{\parallel,tol}$, where $\Delta\Phi_{\parallel,tol}$ is the tolerance of wavefront aberration for the target polarization. Then the units are selected and patterned by minimizing the wavefront aberration for the undesired orthogonal polarization. In contrast to the conventional design principle, where the target phase maps are matched by several pre-selected units, our design is achieved by selecting an arbitrary element with a matched phase for each position.\\
\indent As a proof of concept, we first demonstrate a linear-polarization-sensitive achromatic metalens with an aperture size of 20 $\rm{\mu m}$ and a focal length of 100 $\mu$m. The metalens is implemented by patterning anisotropic $\rm{TiO_2}$ meta-atoms in a square lattice with a fixed period of 400 nm. The heights of the meta-atoms are identical (1 $\mu$m) to ease the fabrication. By varying the cross-sectional shapes and in-plane parameters, a database recording both geometrical parameters and optical responses (the polarization-dependent phase and transmission spectra) is constructed by simulations with the finite difference time domain (FDTD) method. Figure \ref{fig2}a shows the database for three visible wavelengths (450 nm, 550 nm, 650 nm) under x-polarization. The blue dots represent the corresponding spectral phases (i.e., the structural dispersion) of meta-atoms with varying geometrical parameters. For clarity, the projected phases in the $\Phi_{450\rm{nm}}$-$\Phi_{550\rm{nm}}$, $\Phi_{450\rm{nm}}$-$\Phi_{650\rm{nm}}$, and $\Phi_{550\rm{nm}}$-$\Phi_{650\rm{nm}}$ planes are also depicted as red, yellow, and green dots, respectively.
Note that the structural dispersions (blue dots) cannot cover the entire phase space. Therefore, the requirement of spectral phases located in the blank areas cannot be met. To address this issue, the wavelength-dependent reference phase is intelligently adjusted with the assistance of the particle swarm optimization (PSO) algorithm. For example, the required spectral phases (orange dots) for the $\Phi_{550\rm{ nm}}-\Phi_{650 \rm nm}$ plane is gradually shifted into the covering range of our database (blue dots) during the iterations (Figure \ref{fig2}b). Then, meta-atoms offering the required dispersion can be selected to  match better the designed spatial phase profiles for the three wavelengths (Figure \ref{fig2}c,d). The advantage of our design method is more evident when considering more wavelengths (e.g., five wavelengths including 450 nm, 500 nm, 550 nm, 600 nm, and 650 nm) for better broadband achromatic capability. Attributed to the sufficiently small phase errors (Figure \ref{fig2}e), the spectral focal lengths are identical and consistent with the design counterpart (Figure \ref{fig2}f).\\
\indent By rotating the sub-metalens with an angle of $90^\circ$ or $45^\circ$, an optimized sub-metalens targeting the y-polarization or $45^\circ$-polarization can be obtained. Meanwhile, sub-metalens focusing the left-handed circular polarization (LCP) component is designed by incorporating the PB phase and dynamic phase with the assistance of our design method (Supplementary Section S1). 
\section{Results and Discussion}
\subsection{Optical characterizatio}
\begin{figure}[H]
\centering
\includegraphics[width=\textwidth]{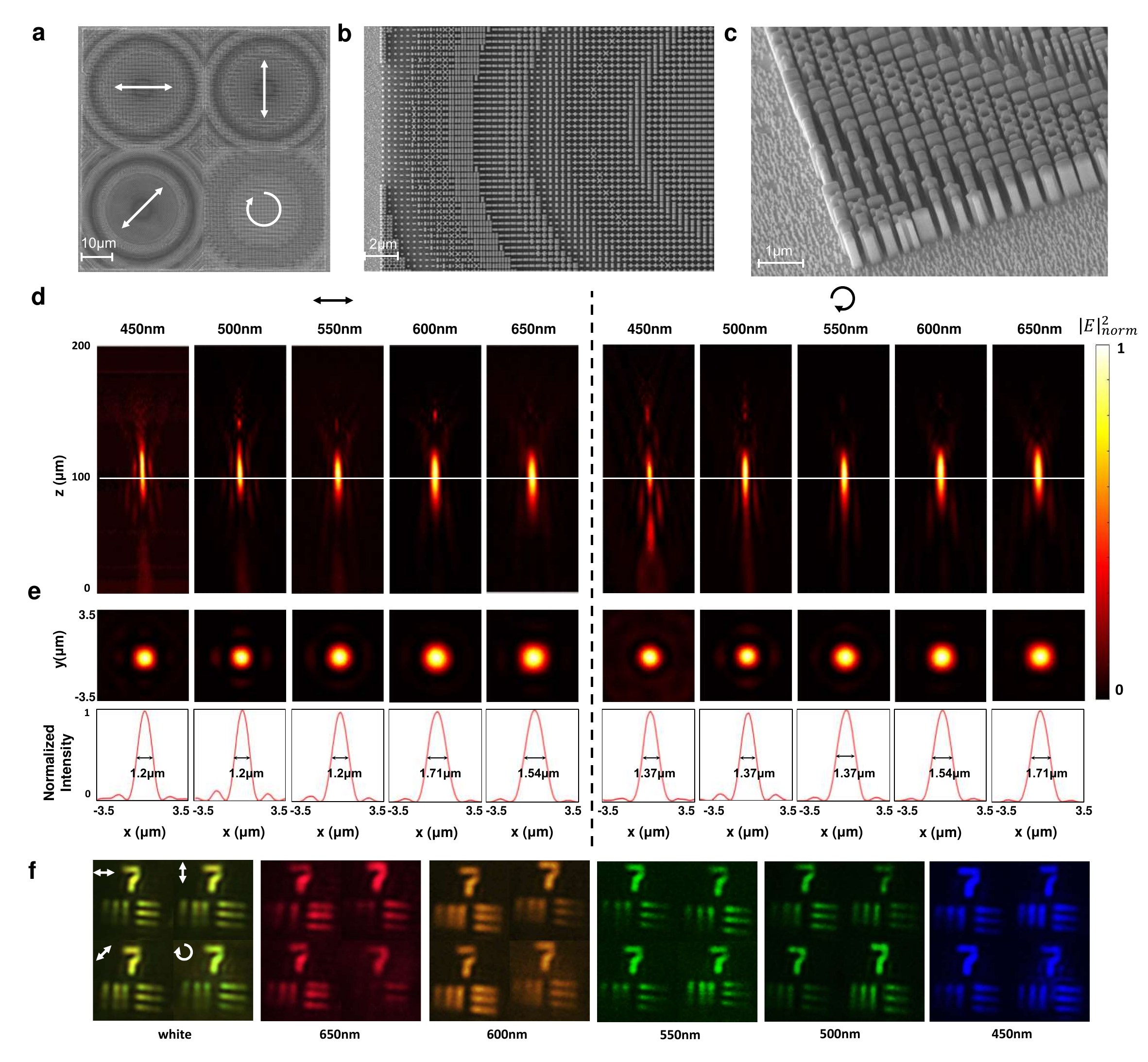}
\caption{(a) Global SEM image of the polarimeter from the top view. (a) Local SEM image of the polarimeter from the top view. (c) Oblique view of the part of the polarimeter. (d) Measured intensity distributions in the x-z plane at sampled incident wavelengths for 450 nm, 500 nm, 550 nm, 600 nm, and  650 nm under x polarized incidence (left) and LCP incidence (right). The white dashed lines are the position of the focal plane. (e) The measured focal plane intensity distributions. (f) Images of the USAF (United States Air Force) resolution target with white light and monochromatic illuminations.}
\label{fig3}
\end{figure}

The broadband achromatic metasurface featuring polarization-spectral imaging is prepared by electron-beam lithography followed by the atomic layer deposition and ion beam etching process (Supplementary Section S3). The fabricated metasurfaces show excellent pattern fidelity consistent with our design (Figure \ref{fig3}a-c). As demonstrations, sub-metalenses' performance for x- and LCP polarization are characterized using a white light source with color filters and polaroids (Supplementary Section S2). The focal lengths of both sub-metalenses are identical and wavelength-insensitive (the white lines in Figure \ref{fig3}d), which is consistent with the theoretical predictions (Figure \ref{fig2}f). The incident waves with different wavelengths are all modulated into diffraction-limited foci (Figure \ref{fig3}e). We further quantified the focusing efficiency of the two sub-metalens which is defined as the ratio of optical power at the focal spot to the total power incident on the sub-metalens aperture. As depicted in Figure S4, the focusing efficiencies for the target polarization surpass $80\%$ for most wavelengths, accompanying negligible crosstalk (peak extinction ratio $\sim20$).
Then the imaging performance of the whole polarimetry is characterized using the USAF (United States Air Force) resolution target. As shown in Figure \ref{fig3}f, element NO. 1 in group NO. 7 (the feature size as 3.9 $\rm{\mu m}$) is well resolved under white light illumination and monochromatic illuminations. The slight blurs result from coma aberration as the resolution target is off-axis to all sub-metalens. The result under white light is yellowish due to the intrinsic color cast by the white light source. 
\subsection{Broadband achromatic full-stokes detection}
\begin{figure}[H]
\centering
\includegraphics[width=\textwidth]{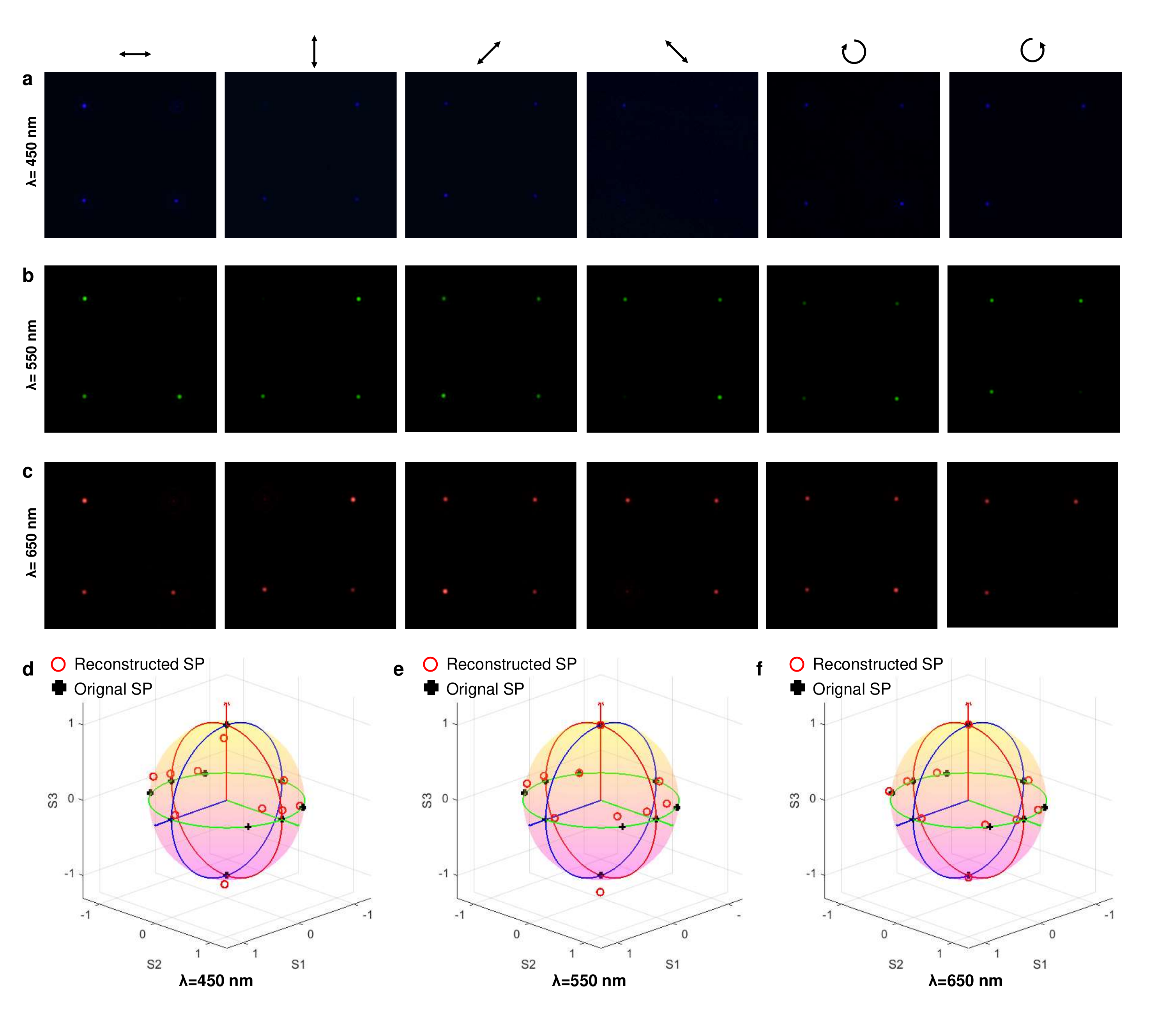}
\caption{Characterization results of the full-color polarimeter in the visible. Measured intensity distributions at the focal plane for different incident polarizations (shown with white arrows) with wavelengths of (a) 450 nm, (b) 550 nm, and (c) 650 nm. Spots with the 450 nm wavelength are colored in white for clarity. Reconstructed Stokes parameters and the original counterparts visualized in the Poincaré sphere for the incident wavelengths of (d) 450 nm, (e) 550 nm, and (f) 650 nm. The black crosses represent the Stokes parameters of the incident light (original SP), while the red rings indicate the calculated Stokes parameters from the measured foci (reconstructed SP). }
\label{fig4}
\end{figure}

To demonstrate the broadband polarization detection capability of the fabricated meta-polarimetry, intensity distributions with different polarizations and wavelengths at the focal plane ($f=$100 $\rm{\mu m}$) are measured. Ten types of polarizations, including the six basic polarizations and other elliptical states, with wavelengths of 450 nm, 550 nm, and 650 nm are selected for the experimental demonstration (Tables S1-S3). As shown in Figure \ref{fig4}a-c, each incident polarization state generates an achromatic unique intensity distribution, indicating that each sub-metalens exhibits high polarization selection characteristics as well as achromatic capability. Then the corresponding full Stokes parameters (SP) are calculated according to Eq. S4, and the Muller matrix of our fabricated metalens is subsequently determined with six basic polarizations (Supplementary Section S5). To visually compare the measured results (reconstructed Stokes parameters) and the actual counterparts (original Stokes parameters), their positions in a Poincaré sphere are presented as shown in Figure \ref{fig4}d-f. The measured polarizations are in good accordance with the original counterparts, with the mean relative errors of the six polarization states $7.5\%$, $5.9\%$, and $3.8\%$ at operating wavelengths of 450 nm, 550 nm, and 650 nm (Tables S1-S3). Therefore, our metalens features an outstanding ability to decompose the different polarization components of the incident light.
\section{Conclusion}
In summary, full-color full-Stokes detection/imaging is enabled by our proposed achromatic polarimeter consisting of four polarization-dependent sub-metalenses. In contrast to the conventional table look-up method which requires a huge database filling all phase spaces, the all-in-one metasurface is designed by a PSO-assisted scheme that matches arbitrary phase compensation with limited database. Such a single-layer metasurface device directly operates with unpolarized beams for full-polarization imaging with broadband achromaticity in the visible. The reconstructed errors of polarization for operating wavelengths of 450 nm, 550 nm, and 650 nm are $7.5\%$, $5.9\%$, and $3.8\%$, respectively. Owing to the broadband achromaticity, high focusing efficiencies ($> 80\%$) and low crosstalk (peak extinction ratio $\sim20$), the spectral and polarization information of objects can be simultaneously revealed when collaborating with a full-color camera. This work would extensively promote integrated devices based on meta-optics.
\begin{acknowledgement}
We acknowledge the financial support by the National Natural Science Foundation of China (Grant Nos. 62275078, 62105120), Natural Science Foundation of Hunan Province of China (Grant No. 2022JJ20020), Basic and Applied Basic Research Foundation of Guangdong Province (Grant No. 2022A1515140113), Youth Innovation Funds of Ji Hua Laboratory (Grant No. X220221XQ220), Science and Technology Innovation Program of Hunan Province (Grant No. 2023RC3101), and Shenzhen Science and Technology Program (Grant No. JCYJ20220530160405013).

\end{acknowledgement}

\begin{suppinfo}

\end{suppinfo}

\bibliography{References}
\section{Supplementary Information}\label{SI}
\subsection{Section S1 Wavefront manipulation for polarization-sensitive focusing}
To achieve direct full Stokes polarimetry, the wavefront for the orthogonal polarization should be manipulated simultaneously to suppress the same focusing. For the linear polarization, we choose another focusing phase for the orthogonal polarization. Therefore, the optimized metalens phase should satisfy two conditions that can be expressed as follows:
\begin{equation}
\left\{
\begin{aligned}
		& \phi_{//}(r,\lambda)=-\dfrac{2\pi}{\lambda}\left(\sqrt{r^2+f_t^2}-\sqrt{r_0^2(\lambda)+f_t^2}\right)\\
		& \phi_{\perp}(r,\lambda)=-\dfrac{2\pi}{\lambda}\left(\sqrt{r^2+f_o^2}-f_o\right)
	\end{aligned}
 \right.
 \tag{S1}
 \label{eqS1}
\end{equation}
where $\phi_{//}(r,\lambda)$ and $\phi_{\perp}(r,\lambda)$  are the phase profiles of the target polarization and the orthogonal polarisation, respectively.\\ 
\indent For the circular polarization, the broadband achromatism is achieved by merging the geometric phase and dynamic phase:
\begin{equation}
\left\{
\begin{aligned}
		& \phi_{//}(r,\lambda)=\Phi_{\lambda_d}(r)+\Delta\Phi_d\left(r,\lambda,\lambda_d\right)\\
		& \phi_{\perp}(r,\lambda)=-\Phi_{\lambda_d}(r)+\Delta\Phi_d\left(r,\lambda,\lambda_d\right)
	\end{aligned}
 \right.
 \tag{S2}
 \label{eqS2}
\end{equation}
where $\Phi_{\lambda_d}=-2\pi/\lambda\left(\sqrt{r^2+f_t^2}-\sqrt{r_0^2(\lambda_d)+f_t^2}\right)$ is achieved by the geometric phase accompanying the polarization conversion; $\Delta\Phi_d\left(r,\lambda,\lambda_d\right)=\Phi_d(r,\lambda)-\Phi_d(r,\lambda_d)$ is the spectrally dynamic phase for the designed wavelength $\lambda_d$. Since a divergent phase for the opposite spin is inherently obtained, the corresponding wavefront optimization could be beyond our consideration.
\subsection{Section S2 Numerical simulations}
The optical responses of the meta-atoms were simulated by the finite-difference time-domain (FDTD) method and the built-in Particle swarm optimization (PSO) algorithm. (Lumerical FDTD Solutions). The period and height of meta-atoms were set as 400 nm and 1 $\mu$m. Considering the constraints of experimental conditions and period size, we set the minimum and maximum size constraints to be 50 nm and 370 nm, respectively. The refractive index of the $\rm TiO_2$ was measured by an ellipsometer. To get the optimal structure parameters with desired polarization-dependent phase and dispersion, the PSO algorithm with a generation size of 10 and maximum generations of 30 was carried out. The termination condition of the iteration was defined as the difference between the current optimal solution being calculated and the average value of the previous three generations is 0.
\begin{figure}[H]
\centering
\includegraphics[width=\textwidth]{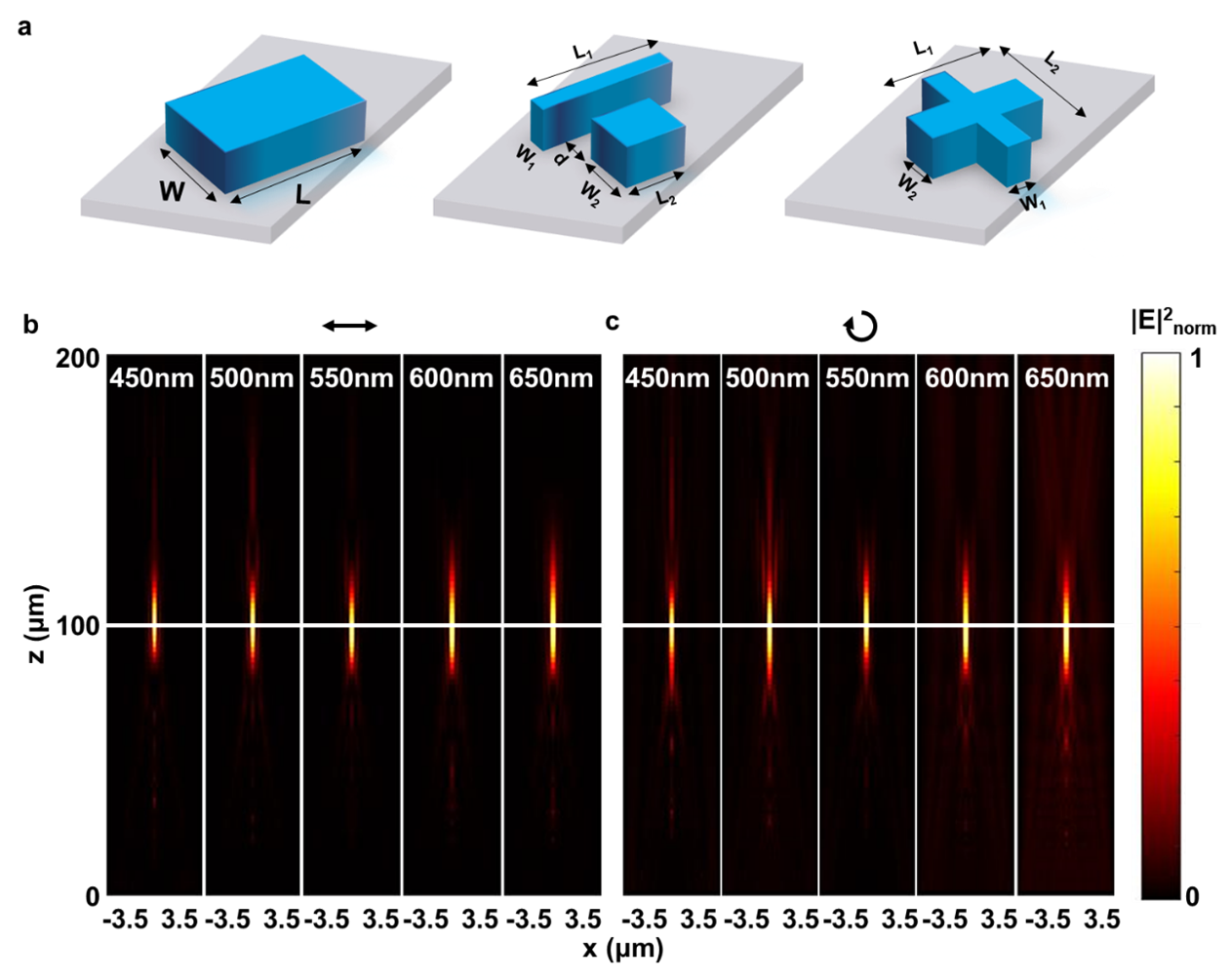}
\label{figS1}
Figure S1 Schematic of the fabrication process of our designed metasurface.
\end{figure}
\subsection{Section S3 Device fabrication} 
The broadband achromatic metasurface featuring polarization-spectral imaging is prepared by electron-beam lithography followed by the atomic layer deposition and ion beam etching process, as depicted in Figure S2.
\begin{figure}[H]
\centering
\includegraphics[width=\textwidth]{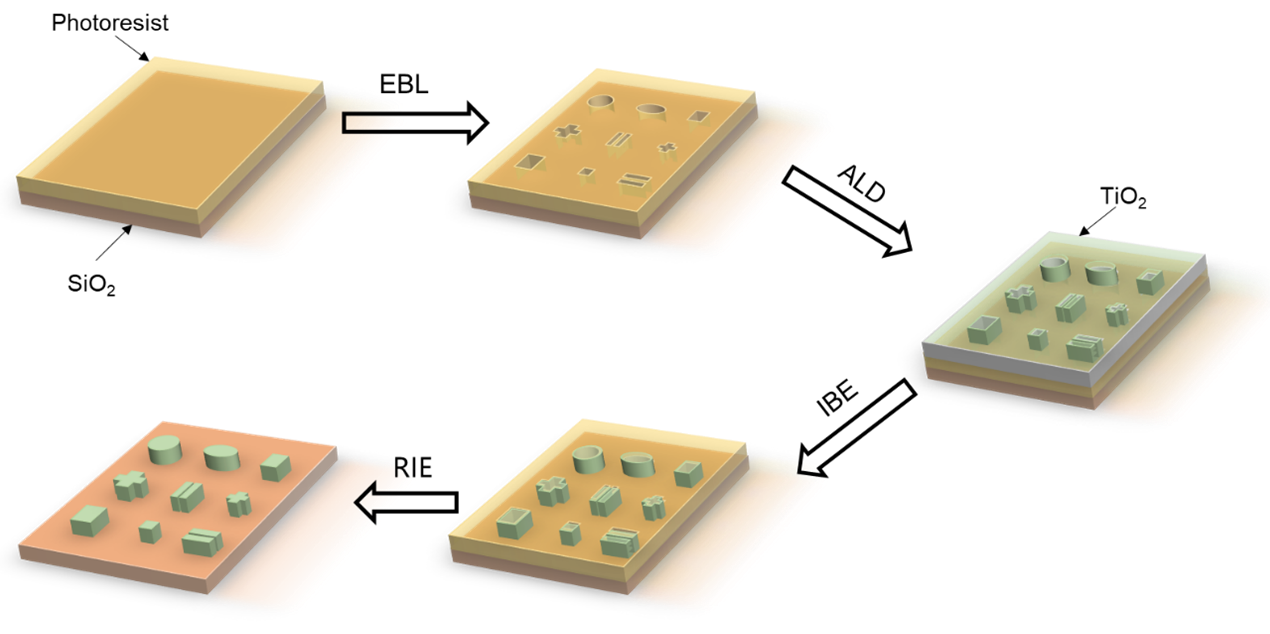}
\label{figS2}
Figure S2 Schematic of the fabrication process of our designed metasurface.
\renewcommand{\thefigure}{Figure S2}
\end{figure}
\subsection{Section S4 Optical measurement for the designed metasurface polarimetry}
The optical setup for the focusing characterization is depicted in Figure S3a. The white light was generated with a supercontinuum laser, and a monochromatic beam with a typical wavelength was achieved by combining a filter. Arbitrarily polarization was then manipulated with a polaroid (LP), and a quarter-wave plate (QWP). The focusing profile was enlarged by a 50X objective lens and then captured by a CCD camera. By tuning the position of the objective lens with an electronically controlled translation stage, the three-dimensional intensity profile was recorded. By replacing the supercontinuum laser with a Halogen lamp, and placing the object in front of the metalens, the optical characterization for polarization imaging was enabled (Figure S3b).
\begin{figure}[H]
\centering
\includegraphics[width=\textwidth]{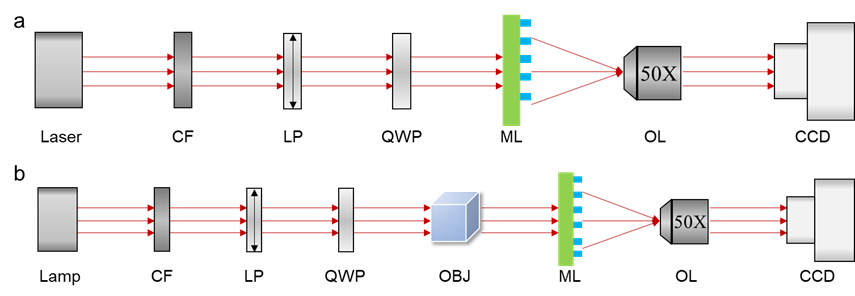}
\label{figS3}
Figure S3 The optical setup for (a) focusing characterization and (b) imaging polarimetry. Optical elements: Color filter (CF), linear polarizer (LP), quarter-wave plate (WP), metalens (ML), objective lens (OL), object (OBJ).
\end{figure}
The focusing efficiencies of the two sub-metalenses were subsequently quantified by calculating the ratio of optical power at the focal spot (concentrated within the triple size of the full width at half maximum) to the total power incident on the sub-metalens aperture. As depicted in Figure S4a, the focusing efficiencies for the target polarization surpass $80\%$ for most wavelengths. In particular, for the linear-polarization-sensitive sub-metalens, the extinction ratio, which is defined as the ratio of the focusing intensity with its target polarization and its orthogonal polarization, reaches nearly 20 at the wavelength of 550 nm (Figure S4b). 
\begin{figure}[H]
\centering
\includegraphics[width=\textwidth]{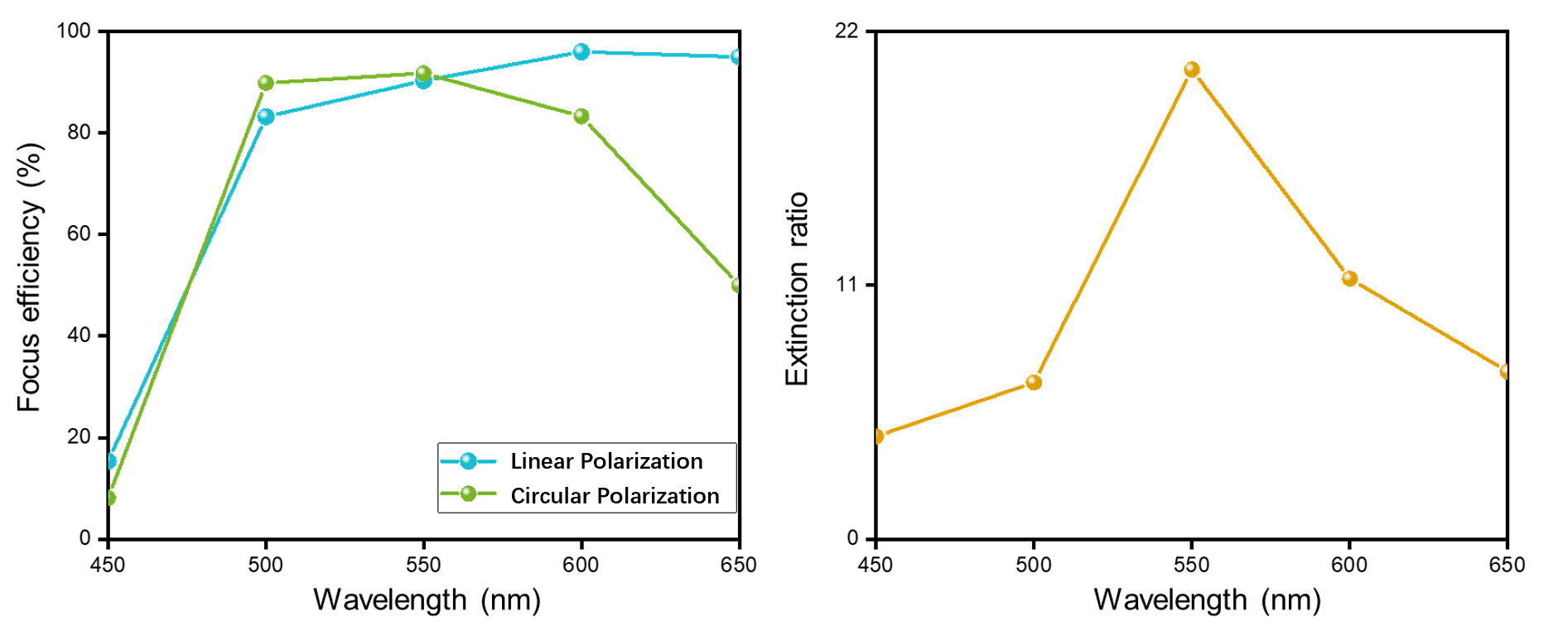}
\label{figS4}
Figure S4 measured focusing efficiency and the extinction ratio of the five wavelengths.
\end{figure}
\subsection{Section S5 Method for Polarization Reconstruction}\label{SI5}
To eliminate the interference of factors such as background light and experimental errors, the measurement is carried out through the Muller Matrix ($\mathbf{M}$) of the metasurfaces: 
\begin{equation}
\mathbf{S=MI_{out}}	
 \tag{S3}
 \label{eqS3}
\end{equation}
Here, $\mathbf{ I_{out}}=[I_{\rm{xp}},I_{\rm{yp}},I_{\rm{45^\circ}},I_{\rm{L}}]^\mathsf{T}$ is the detected intensity of the four foci, and $\mathbf{S}$ is the Stokes vector defined as
\begin{equation}
\mathbf{S}=
\begin{bmatrix}
		S_0\\S_1\\S_2\\S_3\\
	\end{bmatrix}
\begin{bmatrix}
		I_x+I_y\\I_x-I_y\\I_{45^\circ}-I_{-45^\circ}\\I_R-I_L\\
	\end{bmatrix}
 \tag{S4}
 \label{eqS4}
\end{equation}
where $S_0$ describes the total intensity of the incident light, $S_1$, $S_2$, and $S_3$ represent the polarization state of the incident light. $I_x, I_y, I_{45^\circ}, I_{-45^\circ},I_L$, and $I_R$ are the light intensities
of the $x$-, $y$-, $45^\circ$- and $-45^\circ$-linear polarizations, and left-handed, right-handed circular polarization, respectively. The six basic polarizations are used to determine the metasurface’s Muller Matrices at the three wavelengths:
\begin{equation}
\mathbf{M_{650nm}}=
\begin{bmatrix}
		1.395 &1.172 &0.745 &0.983\\2.943&-1.988&-0.440&-0.241\\-2.043&-2.562&4.564&-0.381 \\-2.335&-1.804&0.650&5.960
	\end{bmatrix}
  \tag{S5-1}
 \label{eqS51}
\end{equation}
\begin{equation}
\mathbf{M_{550nm}}=
 \begin{bmatrix}
 1.853&1.898&0.693&0.146\\2.570&-1.822&-0.638 &-0.205\\-1.59 &-2.393 &3.898 &-0.046\\-3.707&-3.797&1.745&4.228
 \end{bmatrix}
 \tag{S5-2}
 \label{eqS52}
 \end{equation}
 \begin{equation}
 \mathbf{M_{450nm}}=
 \begin{bmatrix}
 1.069&1.091 &0.863 &1.044\\2.460 &-1.711 &-1.131 &0.273\\-2.133 &-2.621 &4.725 &-0.588\\-1.346 &-2.708 &0.545 &4.075
 \end{bmatrix}
 \tag{S5-3}
 \label{eqS53}
\end{equation}
Using the corresponding Muller Matrices with the detected intensity, the full-stokes parameters of the objects can be determined. The specific Stokes parameters shown in Figure\ref{fig4}d-f in the main text are listed in Table S1-S3, where small relative errors between the theoretical and measured parameters indicate the strong polarization detection capability of our fabricated metasurface polarimetry.
\begin{table}
\renewcommand\arraystretch{1.5}
\begin{sloppypar}
Table S1: Comparison of the original and reconstructed full-Stokes parameters for the wavelength of 650 nm.
\end{sloppypar}
  \label{tbSl}
  \begin{tabular}{lll}
    \hline
    Full-Stokes parameters of the incidence & Measured full-Stokes parameters & Relative errors\\
    \hline
    (1,1,0,0)	&	(1,0.882,0.031,-0.005)	&0.051\\
(1,0.5,0.866,0)	&	(1,0.502,0.804,0.012)&	0.025\\
(1,0,1,0)	&	(1,0.081,0.95,0.004)&	0.045\\
(1,-0.5,0.866,0)	&	(1,-0.412,0.85,-0.014)&	0.039\\
(1,-1,0,0)	&	(1,-1.067,0.02,-0.004)&0.031\\
(1,-0.5,-0.866,0)	&	(1,-0.471,-1.041,-0.022)&	0.076\\
(1,0,-1,0)	&	(1,0.07,-1.038,0.004)&	0.037\\
(1,0.5,-0.866,0)	&	(1,0.572,-0.865,0.046)	&0.04\\
(1,0,0,1)	&	(1,0.017,0.019,1.0150)&	0.017\\
(1,0,0,-1)	&	(1,0.019,0.018,-1.014)	&0.017\\
 \hline
Average relative error& &	0.038\\
 \hline
  \end{tabular}
\end{table}

\begin{table}
\renewcommand\arraystretch{1.5}
\begin{sloppypar}
 Table S2: Comparison of the original and reconstructed full-Stokes parameters for the wavelength of 550 nm.
 \end{sloppypar}
  \label{tbS2}
  \begin{tabular}{lll}
    \hline
    Full-Stokes parameters of the incidence & Measured full-Stokes parameters & Relative errors\\
    \hline
(1,1,0,0)	&	(1,0.934,0.094,0.028)&	0.063\\
(1,0.5,0.866,0)		&(1,0.405,0.703,0.072)&	0.11\\
(1,0,1,0)	&	(1,0.012,0.844,0.069)&	0.079\\
(1,-0.5,0.866,0)	&	(1,.449,0.743,0.033)	&0.069\\
(1,-1,0,0)	&	(1,-0.969,0.09,0.027)&	0.049\\
(1,-0.5,-0.866,0)	&	(1,-0.486,-0.879,0.013)&	0.014\\
(1,0,-1,0)	&	(1,0.007,-1.027,0.062)&	0.032\\
(1,0.5,-0.866,0)	&	(1,0.541,-0.799,0.157)&	0.088\\
(1,0,0,1)	&	(1,0.01,0.004,1.002)&	0.005\\
(1,0,0,-1)	&	(1,0.008,0.005,-1.218)	&0.077\\
 \hline
Average relative error& &	0.059\\
 \hline\end{tabular}\end{table}
\begin{table}
\renewcommand\arraystretch{1.5}
\begin{sloppypar}
 Table S3: Comparison of the original and reconstructed full-Stokes parameters for the wavelength of 450 nm.
 \end{sloppypar}
  \label{tbS3}
  \begin{tabular}{lll}
    \hline
    Full-Stokes parameters of the incidence & Measured full-Stokes parameters & Relative errors\\
    \hline
(1,1,0,0)		&(1,0.967,0.032,0.063)	&0.043\\
(1,0.5,0.866,0)	&	(1,0.247,0.889,0.184)	&0.153\\
(1,0,1,0)	&	(1,-0.108,0.9,0.071)	&0.093\\
(1,-0.5,0.866,0)	&	(1,-0.535,0.792,-0.007)&	0.039\\
(1,-1,0,0)	&	(1,-0.978,0.06,0.03)&	0.037\\
(1,-0.5,-0.866,0)	&	(1,-0.492,-1.015,0)&	0.053\\
(1,0,-1,0)	&	(1,-0.078,-1.098,0.051)	&0.076\\
(1,0.5,-0.866,0)	&	(1,0.486,-0.843,0.224)	&0.087\\
(1,0,0,1)	&	(1,0.113,0.061,0.871)	&0.101\\
(1,0,0,-1)	&	(1,0.08,0.041,-1.084)	&0.068\\
 \hline
Average relative error& &	0.075\\
 \hline
  \end{tabular}
\end{table}
\end{document}